\documentclass[pra,twocolumn,superscriptaddress,longbibliography]{revtex4-1}

\usepackage{amssymb,amsmath,amsfonts}
\usepackage{graphicx}
\usepackage{color,xcolor}
\usepackage{hyperref}
\usepackage{bm}
\usepackage{multirow}
\usepackage{natbib}

\usepackage{eufrak}
\usepackage{comment}
\DeclareMathOperator{\Tr}{Tr}
\DeclareMathOperator{\re}{Re}
\DeclareMathOperator{\im}{Im}
\DeclareMathOperator{\sgn}{sgn}

\begin{document}

%\definecolor{dkgreen}{rgb}{0,0.6,0}
%\definecolor{gray}{rgb}{0.5,0.5,0.5}
%\definecolor{mauve}{rgb}{0.58,0,0.82}

%\lstset{frame=tb,  	language=Matlab,  	aboveskip=3mm,  	belowskip=3mm,  	showstringspaces=false,columns=flexible,  	basicstyle={\small\ttfamily}numbers=none,  	numberstyle=\tiny\color{gray}, 	keywordstyle=\color{blue},	commentstyle=\color{dkgreen},  	stringstyle=\color{mauve}, 	breaklines=true, 	breakatwhitespace=true  	tabsize=3}

\title{Finite frequency backscattering current noise at a helical edge}
\author{B. V. Pashinsky}

\affiliation{Skolkovo Institute of Science and Technology, 143026 Moscow, Russia}
\affiliation{Moscow Institute for Physics and Technology, 141700 Moscow, Russia Russia}
\affiliation{\hbox{L. D. Landau Institute for Theoretical Physics, acad. Semenova av. 1-a, 142432, Chernogolovka, Russia}}

\author{M. Goldstein}

\affiliation{\hbox{Raymond and Beverly Sackler School of Physics and Astronomy, Tel Aviv University, Tel Aviv 6997801, Israel}}

\author{I. S. Burmistrov}
\affiliation{\hbox{L. D. Landau Institute for Theoretical Physics, acad. Semenova av. 1-a, 142432, Chernogolovka, Russia}}
\affiliation{\hbox{Laboratory for Condensed Matter Physics, National Research University Higher School of Economics, 101000 Moscow, Russia}}

\date{\today}

\begin{abstract}
Magnetic impurities with sufficient anisotropy could account for the observed strong deviation of the edge conductance of 2D topological insulators from the anticipated quantized value. In this work we consider such a helical edge coupled to dilute impurities with an arbitrary spin $S$ and a general form of the exchange matrix. We calculate the backscattering current noise 
at finite frequencies as a function of the
temperature and applied voltage bias. We find that in addition to the Lorentzian resonance at zero frequency, the backscattering current noise features Fano-type resonances at non-zero frequencies. The widths of the resonances are controlled by the spectrum of corresponding Korringa rates. At a fixed frequency the backscattering current noise has non-monotonic behaviour as a function of the bias voltage. 
\end{abstract}

\maketitle

\section{\label{sec:one}Introduction}

The hallmark of two-dimensional (2D) topological insulators is helical edge states~\cite{Hasan2010,Xiao}.  
They exist due to spin-momentum locking caused by the presence of strong spin-orbit coupling \cite{Kane,BHZ}. The helical edge states have been detected experimentally in HgTe/CdTe quantum wells \cite{Konig2007,Roth2009,Gusev2011,Brune2012,Kononov2015}. The time-reversal symmetry protects the helical edge states from elastic backscattering. As a consequence, one expects 
ballistic transport along the helical edge 
with the quantized conductance of $G_0=e^2/h$. However, this idealized picture of edge transport was questioned by experiments 
in a number of 2D topological insulators: HgTe/CdTe quantum wells~\cite{Konig2007,Nowack,Grabecki,Gusev2013,Gusev2014,Kvon2015},  
InAs/GaSb quantum wells~\cite{Knez2011,Suzuki2013,RRDu1,RRDu2,
Suzuki2015,Mueller2015,RRDu3,Mueller2017}, WTe$_{2}$ monolayers~\cite{Cobden17,Jia2017,Herrero2018}, and Bi bilayers~\cite{Sabater2013}.
In order to account for the experimental 
data, several physical mechanisms of backscattering were proposed and studied theoretically, including the effects of electron-electron interaction ~\cite{Xu2006,Schmidt2012,Oreg2012,Altshuler2013,Gornyi2014,Schmidt2015,Yudson2015,Glazman2016,Meir2017,Yudson2018}, charge puddles acting as an effective spin-${1}/{2}$ impurity~\cite{Goldstein2013, Goldstein2014,Nagaev2016,Vinkler-Aviv2020}, a quantum magnetic impurity~\cite{Maciejko2009,Tanaka2011,Cheianov2013,Kimme2016,Kurilovichi2017,Kurilovichi2019a}, the effect of nuclear spins~\cite{Hsu1,Hsu2} etc. 

The average current can provide only limited information on a source of backscattering. More details on scattering can be extracted from current---current correlations, i.e, from current noise \cite{BB,LS,Khrapai1}. To obtain such information one needs to consider current noise beyond the linear response regime, i.e., at finite frequency and/or finite bias voltage. However, so far the shot noise at the helical edge has attracted much less theoretical and  experimental attention in comparison with the average current \cite{Rosenow2013,Khrapai1,Khrapai2,Nagaev2016,VG2017,Mani1,Nagaev2018,Pikulin2018,Kurilovichi2019b}. 

Recently, the zero-frequency shot noise of backscattering current in the case of a magnetic impurity with anisotropic exchange interaction 
for spin $S = \frac{1}{2}$~\cite{VG2017} and spin $S>1/2$~\cite{Kurilovichi2019b} has been computed. It was found that in the case of spin $S = \frac{1}{2}$ the the zero-frequency  backscattering shot noise Fano factor is bounded from above whereas for spins $S>1/2$ the noise can be of arbitrary large magnitude due to bunching of  backscattering pairs of electrons. 

%%%%%%%%%%%%%%%%%%%%%%%%%%%
\begin{figure}[b]
\centerline{\includegraphics[width=0.44\textwidth]{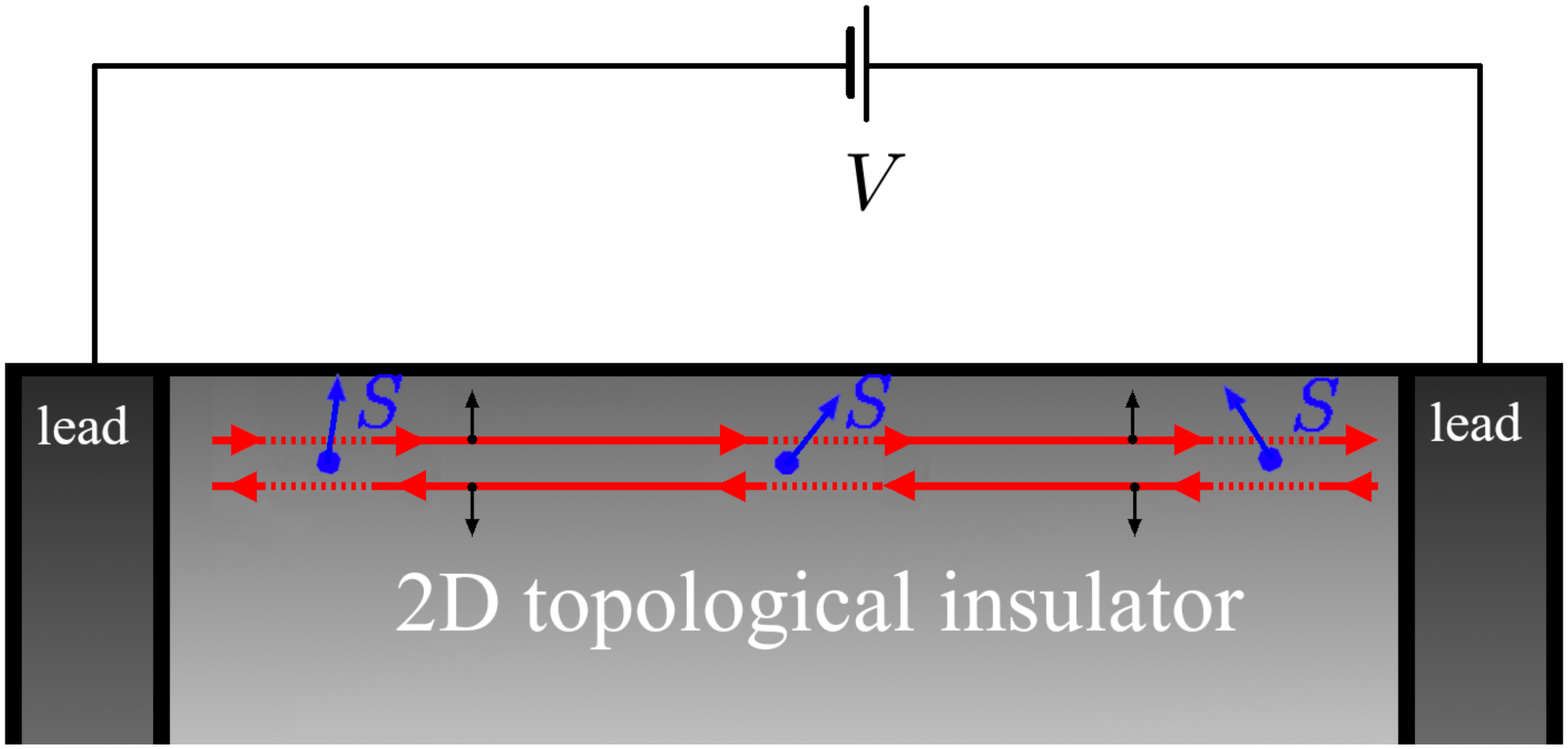}}
\caption{Sketch of the Setup: A helical edge of a 2D topological insulator contaminated by dilute spin-$S$ magnetic impurities. (see text) }
\label{Figure1}
\end{figure}
%%%%%%%%%%%%%%%%%%%%%%%%%

In this paper, we report the results for the finite frequency current noise at the helical edge due to backscattering electrons off a dilute quantum  magnetic impurities with an arbitrary spin $S$ (see Fig. \ref{Figure1}). Similar to Ref. \cite{Kurilovichi2019b} we consider the most general exchange interaction between magnetic impurity and the helical edge states. Under the assumption of weak exchange interaction we derive analytic expression for the current noise as a function of frequency, $\omega$, voltage, $V$, and temperature, $T$. We find that the frequency dependence of the current noise has a resonant structure (see Fig. \ref{Figure3}).
While the resonance at zero frequency has a Lorentzian form, the resonances at non-zero frequencies are of Fano type.
The resonance structure of the current noise is similar to the behavior of the dynamical spin susceptibility under the conditions of electronic paramagnetic resonance. 
In our case voltage plays a role of magnetic field that lifts degeneracy of the impurity spectrum, whereas finite frequency allows transitions between the split energy levels. The broadening of the resonances is determined by the corresponding inverse Korringa times. For $S>1/2$ we find that the peak at zero frequency can become very narrow,  reflecting the aforementioned bunching of backscattered electrons \cite{Kurilovichi2019b}.

The outline of this paper is as follows. We start from describing the formalism in Sec. \ref{sec:two}. In Sec. \ref{sec:three} we present results for the current noise for an arbitrary impurity spin. The special case of spin $S=1/2$ is considered in Sec. \ref{sec:four}. We end the paper with discussions and conclusions in Sec. \ref{sec:five}.

\section{\label{sec:two} Formalism}

\subsection{Model}

The non-interacting 1D helical mode coupled to a magnetic impurity is described by the following Hamiltonian, 
%(there and later $e=-1$,$h=1$):
\begin{equation}\label{s1f1}
H=H_{e}+H_{e-i}.
\end{equation}
Here the first term is the Hamiltonian of edge electrons,
\begin{equation}
H_{e}= iv \int dy\, \Psi^\dag (y) \sigma_{z} \partial_{y}\Psi(y) ,
\end{equation}
where $v$ denotes the velocity of the edge states, $\Psi^\dag$ and $\Psi$ stand for the creation and annihilation operators, and $\sigma_{x,y,z}$ are the standard Pauli matrices operating in the pseudospin space of edge states. The interaction between helical electrons and a magnetic impurity located at $y=y_0$ is assumed to be  in the form of local exchange,
%\cite{Kurilovich1}
\begin{equation}
H_{e-i}=\frac{1}{\nu}\mathcal{J}_{ij}S_i s_j(y_0), \qquad s_j(y) =\frac12\Psi^\dag(y)\sigma_j \Psi(y).
\label{eq:exchange}
\end{equation}
Here $\nu=1/(2\pi v)$ is the density of states per one edge mode and $S_i$ stands for the components of impurity spin operator. 

The $3\times 3$ dimensionless exchange matrix, $\mathcal{J}_{ij}$, $i,j = x,y,z$, is not diagonal due to the presence
of spin-orbit coupling in the 2D topological insulators. For example, there are four nonzero components, $\mathcal{J}_{xx} = \mathcal{J}_{yy}$, $\mathcal{J}_{zz}$, and $\mathcal{J}_{xz}$, for a magnetic impurity in a HgTe/CdTe quantum well in the case of negligible interface
inversion asymmetry \cite{Otten,Kimme2016,Kurilovichi2017}.  
The inversion asymmetry present in HgTe/CdTe quantum wells  \cite{Dai2008,Konig2008,Winkler2012,Weithofer-Recher,Tarasenko2015,Durnev2016,Minkov2013,Minkov2016} renders all matrix elements $\mathcal{J}_{ij}$ nonzero.  
The exchange interaction \eqref{eq:exchange} is expected to be applicable for other 2D topological insulators, e.g., InAs/GaSb quantum wells, WTe$_2$ monolayers, and Bi bilayers. We assume that dimensionless exchange interaction is weak, $|\mathcal{J}_{ij}|\ll 1$. This is fully justified in physical systems.
For example, for Mn$^{2+}$ ion in a HgTe/CdTe quantum well $|\mathcal{J}_{ij}| \sim 10^{-3}$ \cite{Kurilovichi2017a}.

In the Hamiltonian \eqref{s1f1} we neglect the local anisotropy of the impurity spin, described by
$\mathcal{D}_{jk}S_jS_k$. This can be justified for $|\mathcal{D}_{jk}|\ll \max\{\mathcal{J}_{jk}^2 T, |\mathcal{J}_{jk} V|\}$
\cite{Kurilovichi2019a}. 
With neglect of 
the local
anisotropy the exchange matrix $\mathcal{J}_{ij}$
can be brought to a lower triangular form by rotation of the spin basis for the impurity spin $S_i$. We thus assume hereinafter that 
$\mathcal{J}_{xy}=\mathcal{J}_{xz}=\mathcal{J}_{yz}=0$ and $\mathcal{J}_{xx}\mathcal{J}_{yy} > 0$.

In what follows, we shall ignore Kondo-type renormalization of $\mathcal{J}_{jk}$ that is responsible for a logarithmic dependence of exchange interaction on $\max\{T,|V|\}$ \cite{Kurilovichi2017}. 
In addition, we neglect electron-electron interactions along the helical edge in the Hamiltonian \eqref{s1f1}. We discuss the effect of the interactions on the backscattering current noise in Sec. \ref{sec:five}.

\subsection{Backscattering current noise and the generalized master equation}

The presence of a magnetic impurity causes the backscattering of helical states. In the presence of a bias voltage $V$ along the edge, scattering helical states off magnetic impurity produces a backscattering current. 
The spin-momentum locking allows one to relate
the statistics of the total pseudospin projection of the helical states, $\Sigma_z=\int dy\ s_z(y)$, with
the statistics of the number of electrons 
backscattered off the impurity spin during a large time interval $t$,
\begin{equation}
\Delta N(t) =\Sigma_z(t)-\Sigma_z,
\quad \Sigma_z(t)=e^{i H t}\Sigma_z e^{-i H t} .
\end{equation}
Thus cumulant generated function for $\Delta N(t)$ can be written as 
\begin{equation}\label{s1f6}
{G}(\lambda,t)=\ln \Tr \left[e^{i\lambda\Sigma_{z}(t)}e^{-i\lambda\Sigma_{z}(0)}\rho(0)\right].
\end{equation}
Here $\rho(0)$ denotes the initial density matrix of the total system evolving in accordance with the Hamiltonian \eqref{s1f1}. It is convenient to express the cumulant generating function as (see Ref. \cite{Esposito} for a review),
\begin{equation}\label{s1f7}
{G}(\lambda,t)=\ln \Tr \rho^{(\lambda)}(t) .
\end{equation}
Here the generalized density matrix of the system in the presence of counting field $\lambda$ is given as 
\begin{equation}
\begin{split}
\rho^{(\lambda)}(t) & = e^{-i H^{(\lambda)} t}\rho(0) e^{i H^{(-\lambda)} t}, \\ H^{(\lambda)}&=e^{i\lambda\Sigma_{z}/2}H e^{-i\lambda\Sigma_{z}/2} .
\end{split}
\end{equation}
Tracing out the degrees of freedom of helical edge states we can reduce the problem of computation of the cumulant generating function to the evaluation of the reduced generalized density matrix of the magnetic impurity,
\begin{equation}
{G}(\lambda,t)=\ln \Tr_S \rho^{(\lambda)}_S(t) .
\end{equation}
Using the smallness of the exchange interaction, one can derive  
the generalized Gorini-Kossakowski-Sudarshan-
Lindblad (GKSL) equation, which governs the time evolution of $\rho^{(\lambda)}_S(t)$ \cite{Kurilovichi2019b},
\begin{equation}\label{G}
\frac{d\rho_{S}^{(\lambda)}}{dt}=i\left[\rho_{S}^{(\lambda)},H_{e-i}^{\rm mf}\right]
+\eta_{jk}^{(\lambda)}S_{j}\rho_{S}^{(\lambda)}S_{k}
-\frac{\eta_{jk}^{(0)}}{2}\left\{ \rho_{S}^{(\lambda)},S_{k}S_{j}\right\} .
\end{equation}
Here $H_{e-i}^{\rm mf} = \mathcal{J}_{zz}\langle s_z\rangle S_z/\nu$ stands for the mean-field part of $H_{e-i}$ where $\langle s_z\rangle = \nu V/2$ is the average non-equilibrium spin density.
The $3\times 3$ matrix 
\begin{equation}
\eta_{jk}^{(\lambda)}= \pi T (\mathcal{J} \Pi_V^{(\lambda)} \mathcal{J}^T)_{jk}    
\end{equation}
controls the non-unitary evolution of the reduced generalized density matrix. Here we introduced 
\begin{equation}
 \Pi_V^{(\lambda)} = \begin{pmatrix}
 f_\lambda^+\left(V/T\right)
& -i f_\lambda^-\left(V/T\right) & 0 \\
i f_\lambda^-\left(V/T\right)& f_\lambda^+\left(V/T\right)& 0 \\
0 & 0& 1
\end{pmatrix} 
\label{PiV:def}
\end{equation}
and $f_\lambda^\pm(x) = \frac{x}{2}\left(e^{-i\lambda}e^x \pm e^{i\lambda} \right)/\left(e^x-1\right)$.

The average backscattering current can be extracted from the cumulant generating function as 
\begin{equation}
I_{\rm bs} = \lim\limits_{t\to\infty} \frac{d}{dt}\frac{d{G}(\lambda,t)}{d(i\lambda)}\Biggl |_{\lambda=0} .
\label{gen:current}
\end{equation}
The backscattering current noise at a frequency $\omega$ can be also determined with the help of the cumulant generating function \cite{Flindt,Lambert},
\begin{equation}
%\label{s2f45}
\mathcal{S}_{\rm bs}(\omega)=\omega\int\limits_{0}^{\infty}dt\sin(\omega t)\frac{d}{dt}\frac{d^{2}{G}(\lambda,t)}{d(i\lambda)^{2}}\Biggl |_{\lambda=0} .
\label{gen:noise}
\end{equation}
We note that the master equation approach for computation of the current noise is limited to not too high frequencies, $|\omega|\ll \max\{|V|,T\}$ due to the breakdown of the Markovian approximation at short time scales (see, e.g., Ref. \cite{Marcos2010}).

The reduced generalized density matrix $\rho_S^{(\lambda)}(t)$ and, consequently, the cumulant generating function, depend on the choice of the initial reduced density matrix $\rho_S^{(\lambda)}(0)$. Since we are interested in the statistics of the backscattering current in the steady state of the system, we choose the $\rho_S^{(\lambda)}(0)$ to be equal to the steady state density matrix $\rho_S^{\rm st}$, for which the right hand side (r.h.s.) of the GKSL Eq. \eqref{G} equals zero at $\lambda=0$.

\section{\label{sec:three} The backscattering current noise for an arbitrary impurity spin}

\subsection{General expression}

We start analysis of the average backscattering current and noise, cf. Eqs. \eqref{gen:current} and \eqref{gen:noise}, from a derivation of general expressions valid for an arbitrary value $S$ of the impurity spin. As usual, it is convenient to think of the $(2S+1)\times (2S+1)$ reduced density matrix $\rho_S^{(\lambda)}$ as a (super)\-vector $|{\rho}_S^{(\lambda)}\rangle$
of length $(2S+1)^2$. Then the GKSL equation can be rewritten as follows,
\begin{equation}\label{s1f11.1}
\frac{d}{dt}|{\rho}_S^{(\lambda)}\rangle = \bm {\mathcal{L}}^{(\lambda)} |{\rho}_S^{(\lambda)}\rangle ,
\end{equation}
where the $(2S+1)^2\times (2S+1)^2$ matrix $\bm{\mathcal{L}}^{(\lambda)}$ denotes the (super)\-operator corresponding to the r.h.s of Eq. \eqref{G}. For further analysis, it is convenient to introduce the left (super)\-vector $\langle \tilde{0}|$ whose inner product with an arbitrary (super)\-vector $|{\rho}_S^{(\lambda)}\rangle$ gives the trace, $\Tr{\rho}_S^{(\lambda)} \equiv \langle\tilde{0}|{\rho}_S^{(\lambda)}\rangle$.

In order to find the backscattering current noise we need to find $|{\rho}_S^{(\lambda)}(t)\rangle$
to second order in $\lambda$. This can be done by means of perturbation theory in $\lambda$. Let us expand the matrix $\bm {\mathcal{L}}^{(\lambda)}$ in powers of $\lambda$,
\begin{equation}\label{s2f4}
\bm {\mathcal{L}}^{(\lambda)}=\bm {\mathcal{L}}_{0}+ \lambda \bm {\mathcal{L}}_{1}+\lambda^{2}\bm {\mathcal{L}}_2+ \dots .
\end{equation}
The (super)vector $|0\rangle$ corresponding to the steady state density matrix $\rho_S^{\rm st}$ is the right eigenvector for the matrix $\bm {\mathcal{L}}_{0}$ with zero eigenvalue,
\begin{equation}
    \bm {\mathcal{L}}_{0} |0\rangle= 0 .
\end{equation}
We note that $\langle\tilde{0}|$ is the left eigenvector of $\bm {\mathcal{L}}_{0}$ with zero eigenvalue, 
\begin{equation}
    \langle \tilde{0}|\bm {\mathcal{L}}_{0} = 0 .
\end{equation}
Left and right zero eigenvectors of $\bm {\mathcal{L}}_{0}$ satisfy normalization condition $\langle\tilde{0}|0\rangle=1$. 

Solving Eq. \eqref{s1f11.1} perturbatively, we find to the second order in $\lambda$,
\begin{align}\label{G5G6}
    |{\rho}_S^{(\lambda)}(t)\rangle & \simeq|0\rangle +
\int\limits_{0}^{t} dt_1 e^{-\bm{\mathcal{L}}_{0} t_1}\Biggl (\lambda \bm{\mathcal{L}}_{1} +\lambda^2 \bm{\mathcal{L}}_{2} 
\notag \\
& + \lambda^2
 \int\limits_{0}^{t_1}dt_2\ \bm{\mathcal{L}}_{1 }e^{\bm{\mathcal{L}}_{0}(t_1-t_2)}\bm{\mathcal{L}}_{1}
\Biggr )
|0\rangle .
\end{align}
Hence, using the relation ${G}(\lambda,t)=\ln\langle \tilde{0} |{\rho}_S^{(\lambda)}(t)\rangle$, we obtain the series in $\lambda$ expansion of the cumulant generating function,
\begin{align}
  {G}(\lambda,t) = & \lambda \langle \tilde{0} | \Biggl [ t \bigl ( \bm{\mathcal{L}}_{1} +\lambda \bm{\mathcal{L}}_{2}
    \bigr ) + \lambda
    \int\limits_{0}^{t}d\tau (t-\tau)\bm{\mathcal{L}}_{1 } \notag \\
    & \times \Bigl (e^{\bm{\mathcal{L}}_{0}\tau}-|0\rangle\langle \tilde{0} | \Bigr )\bm{\mathcal{L}}_{1} \Biggr ]|0\rangle + O(\lambda^3) .
\end{align}
Next, with the help of Eqs. \eqref{gen:current} and \eqref{gen:noise}, we find the average backscattering current
\begin{equation}
    I_{\rm bs} = -i\langle \tilde{0} | \bm{\mathcal{L}}_{1} |0\rangle  ,
    \label{gen:current:f}
\end{equation}
and the backscattering current noise
\begin{align}
\mathcal{S}_{\rm bs}(\omega) = & 
2\langle \tilde{0}| \bm{\mathcal{L}}_{1}\mathcal{G}\bm{\mathcal{L}}_{1}
-\bm{\mathcal{L}}_{2}
|0\rangle
\notag \\
& -2 \omega^2\langle \tilde{0}| \bm{\mathcal{L}}_{1}\mathcal{G}\left(\bm{\mathcal{L}}_{0}^{2}+\omega^{2}\bm{1}\right)^{-1}\bm{\mathcal{L}}_{1}|0\rangle .
\label{gen:noise:f}
\end{align}
We note that the first line in Eq. \eqref{gen:noise:f} describes the zero frequency noise, while the second line is the frequency dependent contribution. The matrix $\mathcal{G}$ is the pseudo inverse of $\bm{\mathcal{L}}_{0}$. One needs to work with the pseudo inverse matrix since $\bm{\mathcal{L}}_{0}$ has zero eigenvalue. The pseudo inverse $\mathcal{G}$ satisfies the following relation
\begin{equation}\label{L1}
\mathcal{G}\bm{\mathcal{L}}_{0} = \bm{1}-|0\rangle\langle \tilde{0} | .
\end{equation}
With the help of this relation, Eq. \eqref{gen:noise:f} can be rewritten as
\begin{equation}
\mathcal{S}_{\rm bs}(\omega) =  
2\langle \tilde{0}|\Bigl [  \bm{\mathcal{L}}_{1}
\bm{\mathcal{L}}_{0}
\left(\bm{\mathcal{L}}_{0}^{2}+\omega^{2}\bm{1}\right)^{-1} \bm{\mathcal{L}}_{1}
-\bm{\mathcal{L}}_{2}\Bigr ]
|0\rangle
.
\label{gen:noise:f2}
\end{equation}
It is worthwhile to mention that the zero frequence noise is obtained from Eq. \eqref{gen:noise:f2} as $\mathcal{S}(\omega\to 0)$.

In order to characterize the dependence of the backscattering current noise \eqref{gen:noise:f2}
 on the frequency, it is useful to introduce the set of left and right eigenvectors of $\bm{\mathcal{L}}_{0}$ with non-zero eigenvalues $l_\alpha$, $\alpha=1,\dots,4S(S+1)$,
\begin{equation}
    \bm{\mathcal{L}}_{0} |\alpha
\rangle = l_\alpha |\alpha
\rangle, \qquad \langle \tilde{\alpha} |\bm{\mathcal{L}}_{0}  = \langle \tilde{\alpha} | l_\alpha  .
\end{equation}
These eigenvectors are assumed to be mutually orthogonal, $\langle \tilde{0}|\alpha\rangle=\langle\tilde\alpha|0\rangle=0$ and
$\langle\tilde\alpha|\beta\rangle=\delta_{\alpha\beta}$. An eigenvalue $l_\alpha$ can be either pure real or belong to a complex conjugate pair. The real parts of all eigenvalues are negative, $l_\alpha^\prime=\re l_\alpha<0$. Using the system of the eigenvectors, Eq. \eqref{gen:noise:f2} can be written as follows,
\begin{equation}
    \mathcal{S}_{\rm bs}(\omega)  =
    2\sum_{\alpha=1}^{4S(S+1)}
    \frac{l_\alpha \langle \tilde{0}|\bm{\mathcal{L}}_{1}|
   \alpha\rangle \langle \tilde{\alpha}|\bm{\mathcal{L}}_{1}|
   0\rangle}{l_\alpha^2+\omega^2} -  2\langle \tilde 0|\bm{\mathcal{L}}_{2} |0\rangle .
   \label{gen:noise:f3} 
\end{equation}
In the sum over $\alpha$ in the above equation the eigenvalues with zero imaginary part, $l_\alpha^{\prime\prime}=\im l_\alpha=0$,   contribute to the Lorentzian resonance at $\omega=0$. If an  eigenvalue has non-zero imaginary part, it contributes to the resonances at $\omega=\pm l_\alpha^{\prime\prime}$. The form of these resonances are of the Fano type and can be described as, $[c_\alpha^\prime l_\alpha^{\prime 2}+c_\alpha^{\prime\prime}(l_\alpha^{\prime\prime}\pm\omega)]/[l_\alpha^{\prime 2}+(l_\alpha^{\prime\prime}\pm\omega)^2]$. Here we define $c_\alpha^{\prime}+ic_\alpha^{\prime\prime}=\langle \tilde{0}|\bm{\mathcal{L}}_{1}|
   \alpha\rangle \langle \tilde{\alpha}|\bm{\mathcal{L}}_{1}|
   0\rangle$. The width of the resonance is proportional to $|l_\alpha^\prime|$. 

As follows from the GKSL equation \eqref{G}, the real part of $l_\alpha$ is of the order of $J^2 \max\{|V|,T\}$, whereas the imaginary part is of the order of $J V$. Here $J$ is an absolute value of a typical value of the exchange matrix elements $\mathcal{J}_{jk}$. Therefore, at a small bias voltage, $|V|\ll J T$, the backscattering noise has a single Lorentzian resonance at zero frequency with a width of the order of $J^2 T$. At larger voltage, $J T\ll |V|\ll T$, additional side resonances in $\mathcal{S}_{\rm bs}$ become resolved since their widths are still of the order of $J^2 T$. In the case of large voltage, $|V|\gg T$, there are a number of resonances at frequencies of the order of $J |V|$. The widths of these resonances are proportional to $J^2 |V|$. Below we shall discuss this evolution of the backscattering noise with increasing voltage in more detail.

\subsection{The backscattering current noise for $|V|\ll J T$}

As is well-known, the fluctuation dissipation theorem relates the current noise at zero voltage with the linear conductance. Below we shall demonstrate that this relation holds for the backscattering current and noise.

We start from expansion of the matrix $\Pi_V^{(\lambda)}$, Eq. \eqref{PiV:def}, at $V=0$ in series to the second order in $\lambda$. Then from Eq. \eqref{gen:noise:f2} we find for the current noise at $V=0$,
\begin{align}
\mathcal{S}_{\rm bs}(\omega) = & 
\pi T (J \pi_2 J^T)_{jk}\Tr [S_j \rho_S^{\rm st} S_k] 
\notag \\
& + \pi T \sum_{\sigma=\pm}
(J \pi_1 J^T)_{jk} \Tr [S_j \delta \rho_\sigma S_k] .
\end{align}
Here we introduce two matrices 
\begin{equation}
\pi_1=\begin{pmatrix}
0 & -1 & 0\\
1 & 0 & 0 \\
0 & 0 & 0
\end{pmatrix}, \qquad
\pi_2=\begin{pmatrix}
1 & 0 & 0\\
0 & 1 & 0 \\
0 & 0 & 0
\end{pmatrix}.
\end{equation}
Then $\delta \rho_\sigma$ satisfies the following equation,
\begin{gather}
\pi T (JJ^T)_{jk} [S_j \delta\rho_\sigma S_k-\{\delta\rho_\sigma,S_kS_j\}/2]+i \omega\sigma \delta\rho_\sigma
\notag \\
=
\pi T (J \pi_1 J^T)_{jk} S_j \rho_S^{\rm st} S_k .
\label{eq:deltarho:sigma}
\end{gather}
Using the explicit form of matrices $\pi_{1,2}$, we obtain
\begin{equation}
\mathcal{S}_{\rm bs}(\omega) = 
\frac{\pi T S(S+1)}{6} (J \pi_2 J^T)_{jj}
 + \frac{i\pi T}{2} \mathcal{X}_j \sum_{\sigma=\pm}
\Tr [S_j \delta \rho_\sigma] ,
\end{equation}
where $\mathcal{X}_j=2\epsilon_{jkl}\mathcal{J}_{kx}\mathcal{J}_{ly}$.
Solving Eq. \eqref{eq:deltarho:sigma}, we find
\begin{gather}
    \Tr [S_j \delta \rho_\sigma]
    =\frac{S(S+1)}{3i}
    (\Gamma_\sigma(\omega))^{-1}_{jk}\mathcal{X}_k ,
\end{gather}
where we introduce the matrix
\begin{equation}
    (\Gamma_\sigma)_{jk}(\omega) = (\mathcal{J}\mathcal{J}^T)_{jk} +\frac{2 i\sigma \omega}{\pi T} \delta_{jk} - \delta_{jk} (\mathcal{J}\mathcal{J}^T)_{ll} .
    \label{eq:Gamma:s:def}
\end{equation}

Now we perform the singular value decomposition of the exchange matrix $\mathcal{J}={R}_{<}{\Lambda}{R}_{>}$. Here $R_{<,>}$ are the SO(3) matrices and $\Lambda = {\rm diag}\{\lambda_1, \lambda_2, \lambda_3\}$. Combining the results above, we finally obtain the following expression for the backscattering current at $V=0$
\begin{equation}
    \mathcal{S}_{\rm bs}(\omega)=\frac{\pi TS(S+1)}{3}\left(R_{>}^{-1}{\Phi}(\omega) R_{>}\right)_{zz} ,
    \label{noise:low}
\end{equation}
where the matrix ${\Phi}(\omega)$ is given as
\begin{align}
    {\Phi}(\omega) = & \text{diag}\Biggl\{\left(\lambda_{2}^{2}+\lambda_{3}^{2}\right)\frac{\left(\lambda_{2}^{2}-\lambda_{3}^{2}\right)^{2}+4\omega^2/(\pi T)^{2}}{\left(\lambda_{2}^{2}+\lambda_{3}^{2}\right)^{2}+4\omega^2/(\pi T)^{2}},\notag \\
    & \left(\lambda_{1}^{2}+\lambda_{3}^{2}\right)\frac{\left(\lambda_{1}^{2}-\lambda_{3}^{2}\right)^{2}+4\omega^2/(\pi T)^{2}}{\left(\lambda_{1}^{2}+\lambda_{3}^{2}\right)^{2}+4\omega^2/(\pi T)^{2}},\notag \\
   & \left(\lambda_{1}^{2}+\lambda_{2}^{2}\right)\frac{\left(\lambda_{1}^{2}-\lambda_{2}^{2}\right)^{2}+4\omega^2/(\pi T)^{2}}{\left(\lambda_{1}^{2}+\lambda_{2}^{2}\right)^{2}+4\omega^2/(\pi T)^{2}}\Biggr\}.
\end{align}

It is instructive to compare the result \eqref{noise:low} with the result for the average backscattering current at $V\to 0$ \cite{Kurilovichi2017},
\begin{equation}
    I_{\rm bs}=
    \Delta G V, \quad \Delta G=
    -\frac{\pi S(S+1)}{6}\left(R_{>}^{-1}{\Phi}(0)R_{>}\right)_{zz} .
    \label{eq:Ibs:V0}
\end{equation}
Then we find the Nyquist-type relation for the backscattering current noise at zero frequency and voltage,
\begin{equation}
    \mathcal{S}_{\rm bs}(\omega = 0)=2 T |\Delta G| .
    \label{eq:Nyquist:0}
\end{equation}
We mention that the  Nyquist-type relation \eqref{eq:Nyquist:0} implies the absence of correlations between the total current $I$ and the backscattering current $I_{\rm bs}$ in the equilibrium. Indeed, the incoming current $I_{\rm in}=I-I_{\rm bs}$ from reservoirs splits into the transmitted $I$ and backscattered $I_{\rm bs}$ currents. Under equilibrium conditions, the incoming current carries thermal noise, $\mathcal{S}_{in} = 2TG_0$, corresponding to the ideal transport channel. The noise of the total current should obey the fluctuation-dissipation theorem, i.e., is given by  $\mathcal{S}_I=2T (G_0+\Delta G)$. Hence, using Eq. \eqref{eq:Nyquist:0}, we find that the cross correlation of the total current $I$ and the backscattering current $I_{\rm bs}$ vanishes, $\mathcal{S}_{\rm cross} \propto \langle\langle I_{\rm bs} I\rangle\rangle = 0$.

\begin{figure}[t]
\centerline{\includegraphics[width=0.95\columnwidth]{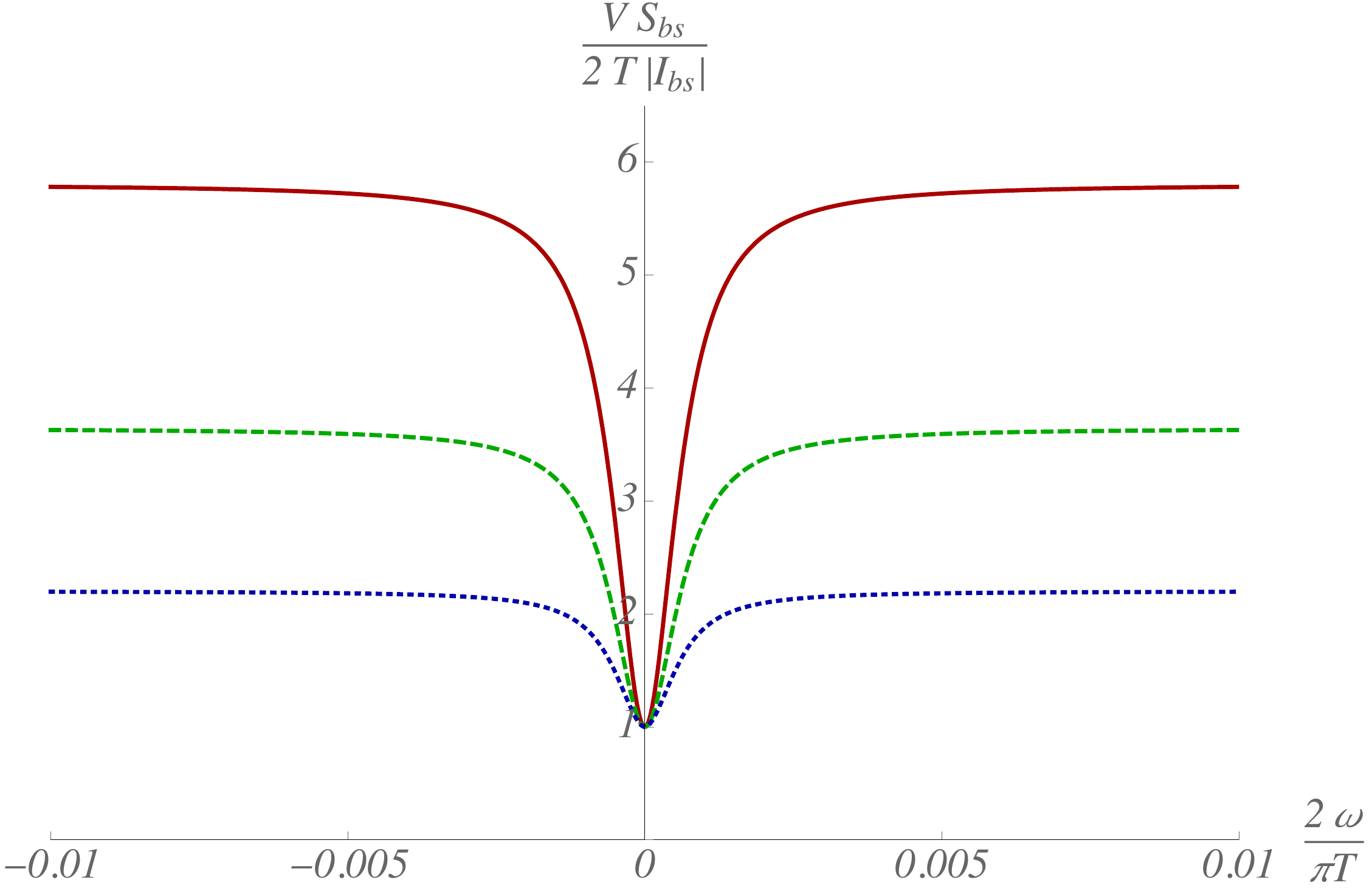}}
\caption{The dependence of the normalized absolute value of backscattering admittance, $V\mathcal{S}_{\rm bs}/(2T |I_{\rm bs}|)$ at $V\ll JT$, on the dimensionless frequency $2\omega/(\pi T)$. The red 
solid curve is plotted for 
$\mathcal{J}_{xx}=\mathcal{J}_{zx}=\mathcal{J}_{yy}/2=\mathcal{J}_{zz}$. The green dashed curve corresponds to $\mathcal{J}_{xx}=\mathcal{J}_{zx}=\mathcal{J}_{yy}/2=2\mathcal{J}_{yx}=5\mathcal{J}_{zy}=\mathcal{J}_{zz}$. The blue dotted curve is plotted for $\mathcal{J}_{xx}/2=\mathcal{J}_{zx}=\mathcal{J}_{yy}=\mathcal{J}_{zz}$. For all curves $\mathcal{J}_{zz}$ is equal to $0.01$.}
\label{Figure2}
\end{figure}

Extending the arguments above to the case of the finite frequency but still zero bias voltage, we obtain that the right hand side of Eq. \eqref{noise:low} determines the absolute value of the backscattering admittance at the helical edge. The dependence of the  backscattering admittance on the frequency is depicted in Fig. \ref{Figure2}. Irs absolute value grows with increasing the frequency. At $|\omega| \sim J^2 T$ the admittance crosses over into a constant in agreement with Eq. \eqref{noise:low}.
\color{black}

%%%%%%%%%%%%%%%%%%%%%%%%%%%
%%%%%%%%%%%%%%%%%%%%%%%%%

\subsection{\label{largefreq}The backscattering current noise for $|V|\gg T$}

Now we consider the limit of large voltage $|V|\gg T$. In this regime the backscatteing current noise determines the full current noise $\mathcal{S}_I$  \cite{VG2017,Nagaev2018}).
\color{black}

At large voltage $|V|\gg T$ the matrix \eqref{PiV:def} simplifies to
\begin{equation}
\Pi_V^{(\lambda)} \longrightarrow 
\frac{V}{2T} e^{- i\lambda \sgn V}
\begin{pmatrix}
1 & -i & 0\\
i & 1 & 0\\
0& 0& 0
\end{pmatrix}.
\label{eq:Pi2}
\end{equation}
Therefore, both 
the backscattering current noise and the backscattering current
are proportional to the voltage and are independent of temperature. The ratio $\mathcal{S}_{\rm bs}(\omega)/|I_{\rm bs}|$ is a function of the dimensionless parameter $\omega/V$. At zero frequency the ratio coincides with the backscattering Fano factor $F_{\rm bs}=\mathcal{S}_{\rm bs}(0)/|I_{\rm bs}|$. This Fano factor  is bounded from below by unity, $F_{\rm bs}\geqslant 1$ \cite{Kurilovichi2019b}. For $S=1/2$ the backscattering Fano factor is also bounded from above, $F_{\rm bs}\leqslant 2$ \cite{VG2017}. For $S>1/2$, $F_{\rm bs}$ is unbounded from above due to bunching of electrons backscattered off a magnetic impurity \cite{Kurilovichi2019b}. 

\begin{figure*}[t]
\centerline{(a)\includegraphics[width=0.45\textwidth]{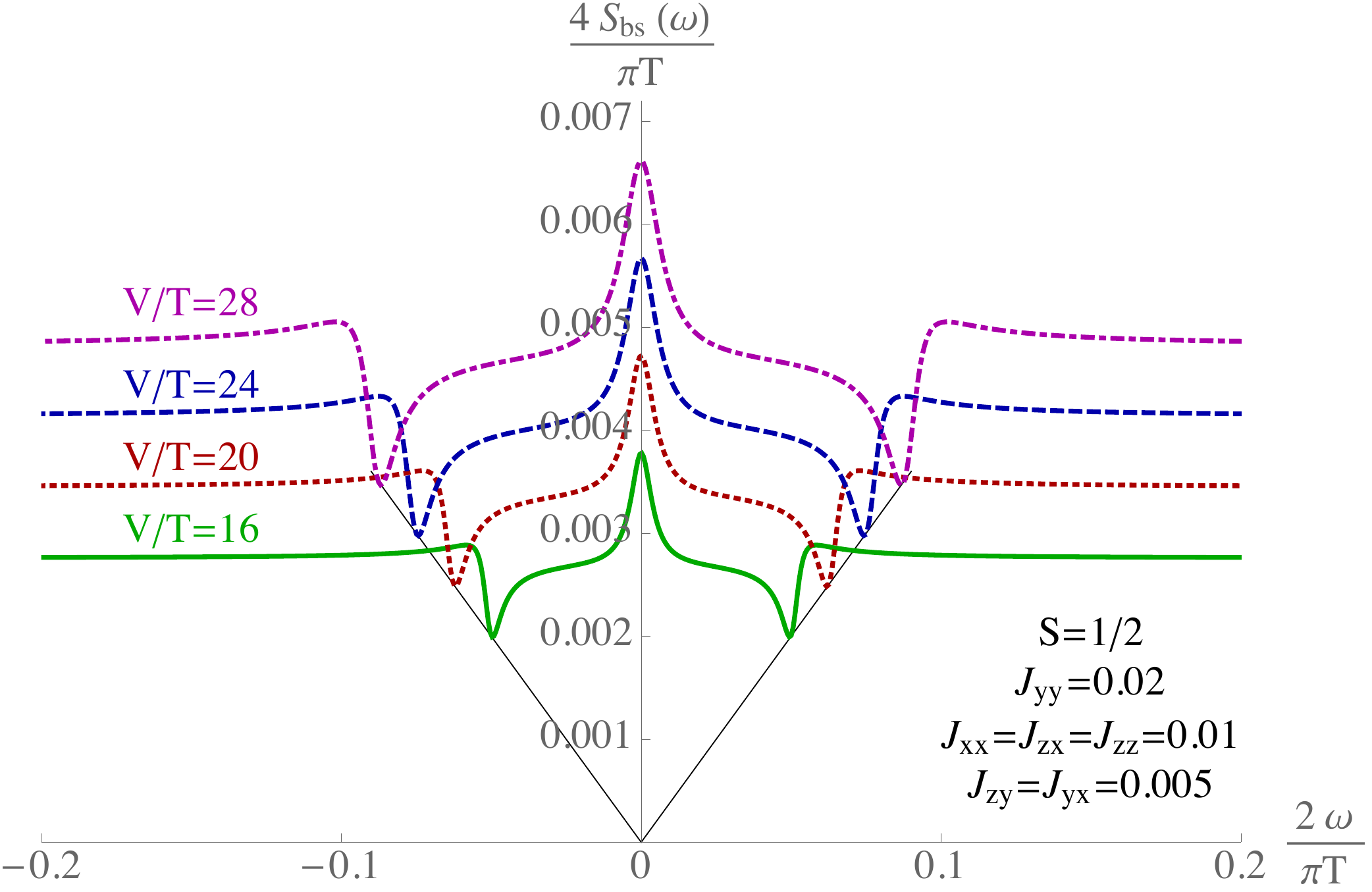}\quad (b) \includegraphics[width=0.45\textwidth]{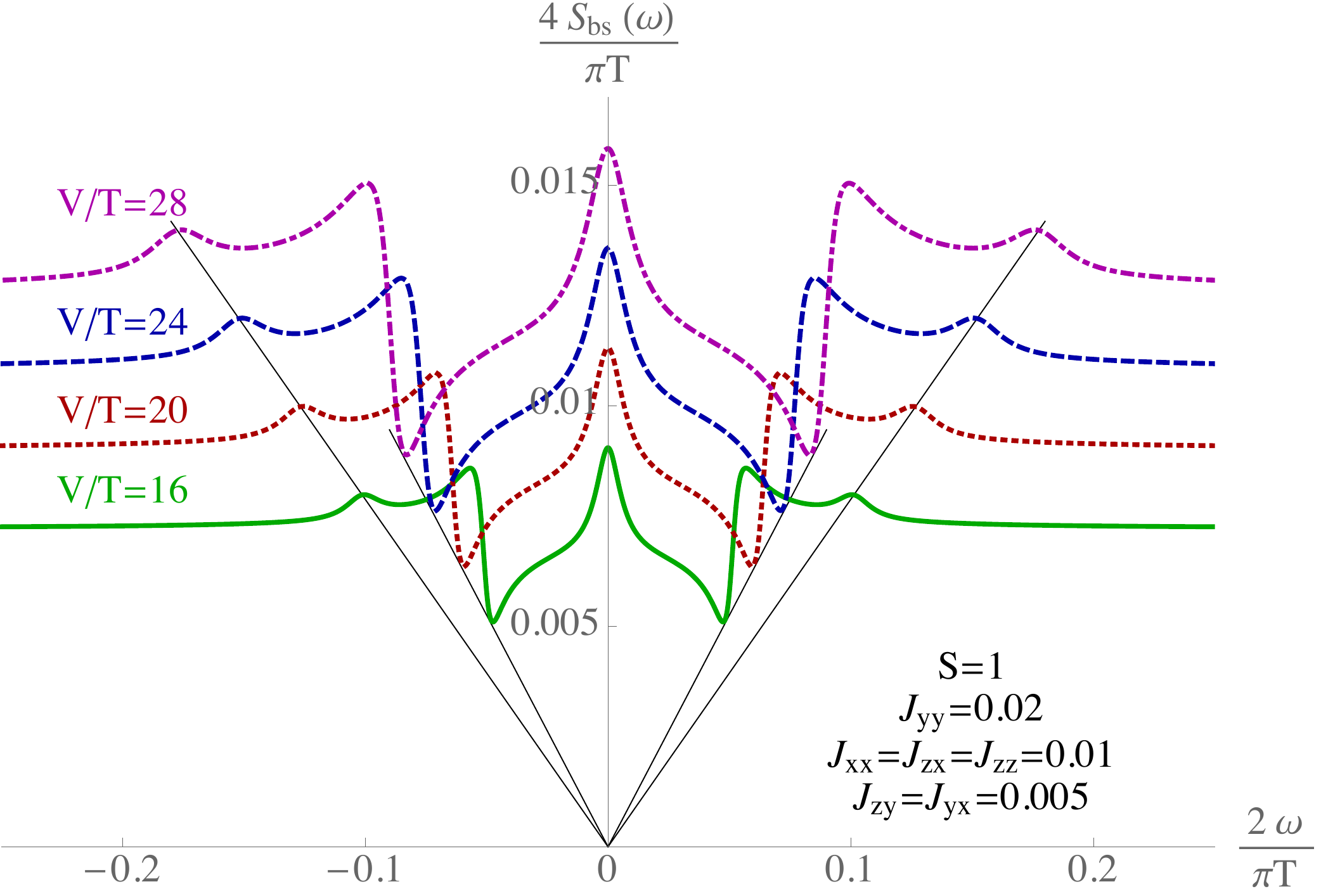}}  
\vspace{0.1cm}
\centerline{
(c) \includegraphics[width=0.45\textwidth]{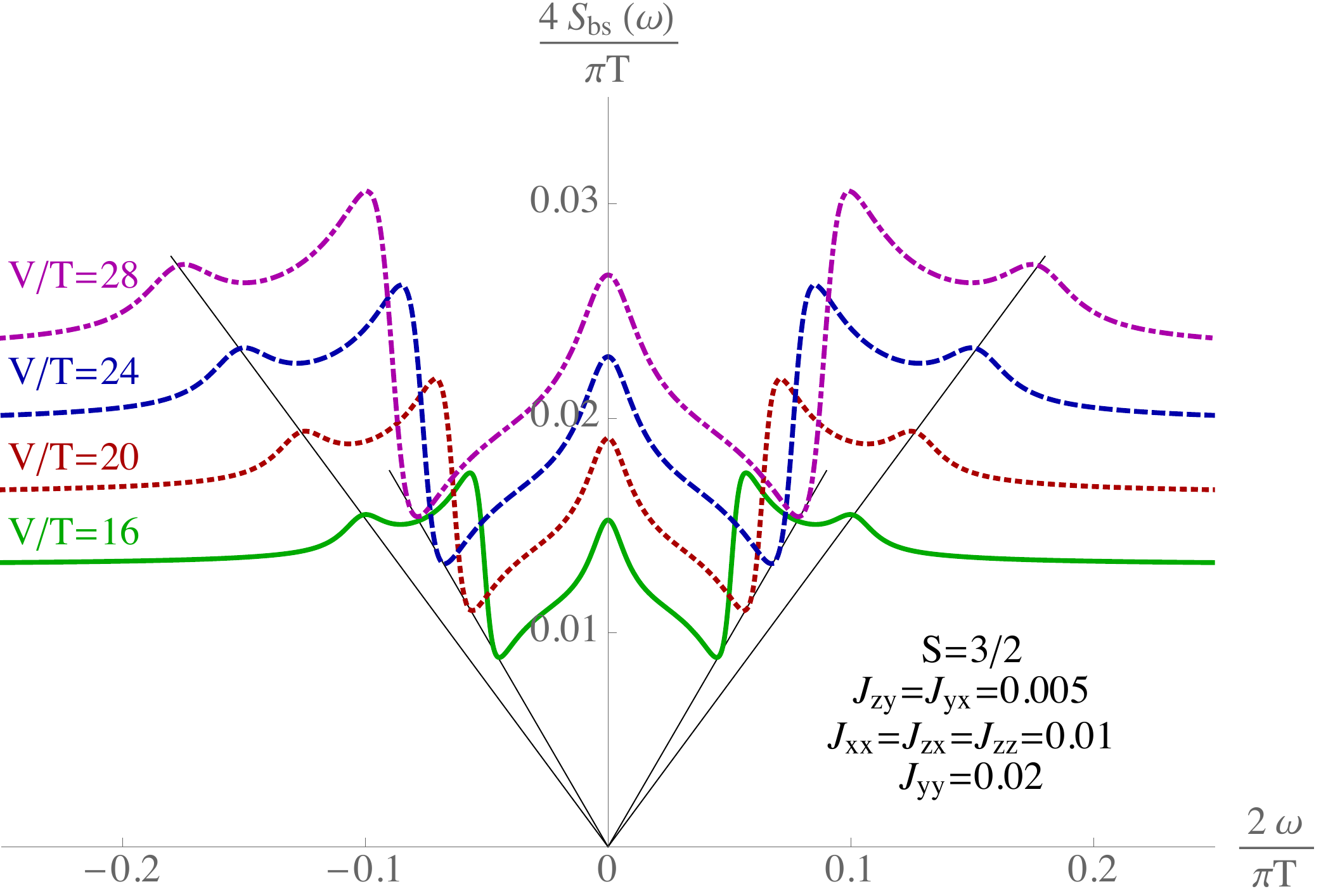}\quad (c) \includegraphics[width=0.45\textwidth]{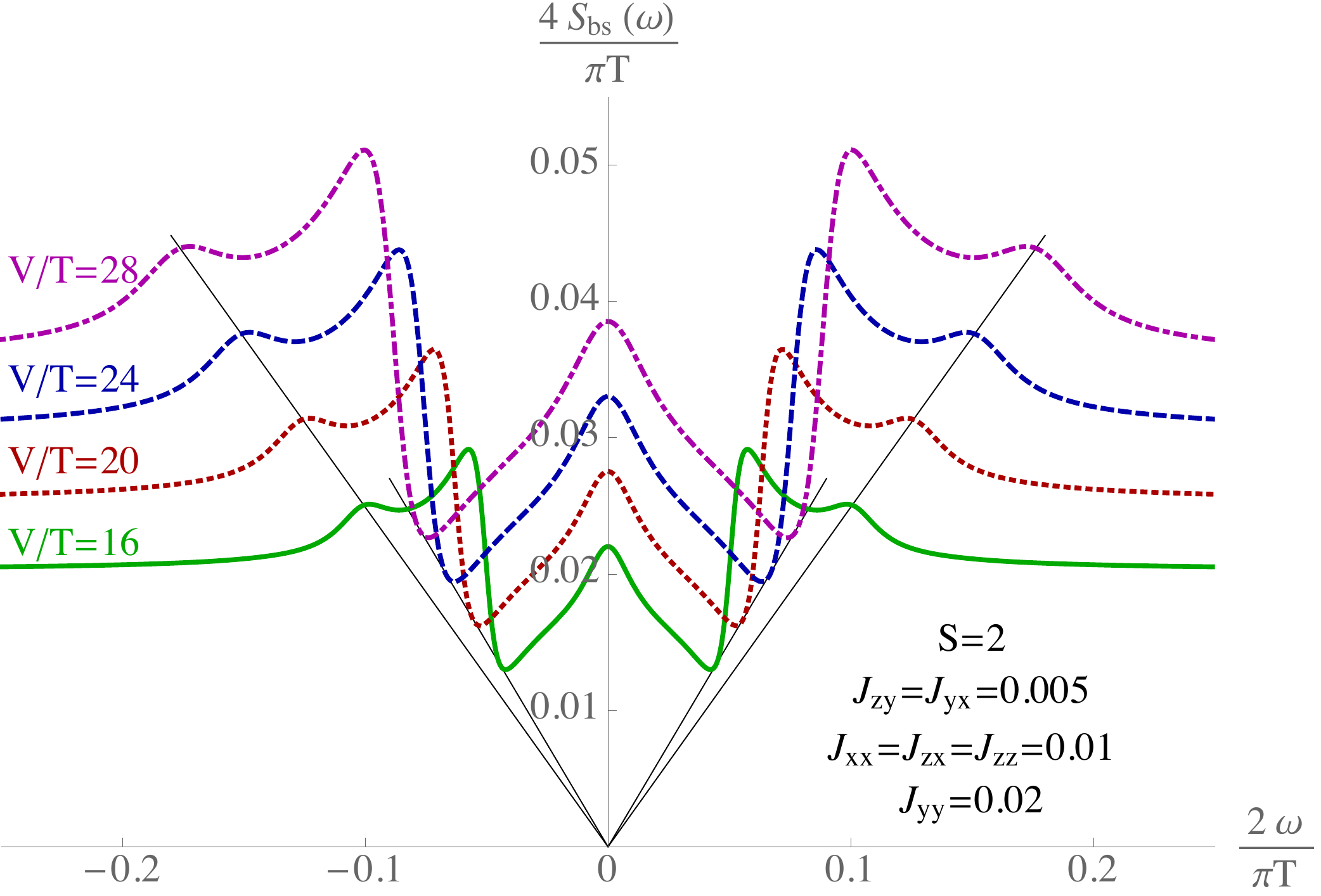}}
\caption{The dependence of the backscattering current noise, $4\mathcal{S}_{\rm bs}/(\pi T)$, on the dimensionless frequency $2\omega/(\pi T)$ at different values of $V/T\gg 1$ and $S$. The black lines are guides for an eye, marking the positions of the resonances.}
\label{Figure3}
\end{figure*}

The behavior of the backscattering current noise as a function of frequency at voltage bias $|V|\gg T$ is shown in Fig. \ref{Figure3} for several magnitudes of the spin of the magnetic impurity. For $S=1/2$, $\mathcal{S}_{\rm bs}(\omega)$ has a maximum at $\omega=0$ and minima at $\omega =\pm J_{zz}V/2$. At larger frequencies the backscattering current noise tends to a constant. In the case $S=1$ the resonances at the frequencies $\pm J_{zz}V/2$ acquires a Fano--like shape and additional maxima appear at $\pm J_{zz}V$. The frequency dependence of the backscattering current noise for $S=3/2$ and $S=2$ resembles the one for $S=1$. Additional resonances which are possible in the case $S=3/2$ and $S=2$ are not visible since their amplitude is of higher order in the small anisotropic part of the exchange interaction matrix $\mathcal{J}$.    

At larger frequencies, $J |V| \ll |\omega| \ll |V|$, $\mathcal{S}_{\rm bs}$  is fully determined by the backscattering current. Indeed, the form 
of the matrix $\Pi_V^{(\lambda)}$ at $|V|\gg T$,
cf. Eq. \eqref{eq:Pi2}, implies the the following relation,
$\mathcal{L}_{2} = -i\sgn V \mathcal{L}_{1}/2$.
Hence, in the limit of large frequencies, $J |V| \ll |\omega| \ll |V|$, we find that the backscattering current noise coincides with the backscattering current, $\mathcal{S}_{\rm bs}(\omega)/ |I_{\rm bs}|\to 1$. This implies that backscattering is completely uncorrelated. We note that a similar result has been derived in Ref. \cite{Nagaev2018} for the case of $S=1/2$. It is worthwhile to mention that depending on the parameters of the exchange matrix $\mathcal{J}_{jk}$, the ratio $\mathcal{S}_{\rm bs}(\omega)/ |I_{\rm bs}|$ can be even smaller than unity at intermediate frequencies, as illustrated in Fig. \ref{Figure3}.

%%%%%%%%%%%%%%%%%%%%%%%%%%%
%%%%%%%%%%%%%%%%%%%%%%%%%

\subsection{The Korringa relaxation rates}

The imaginary parts of eigenvalues $l_\alpha$ determine the position of the Fano-type resonances. These resonances are due to the transitions between levels of the mean field Hamiltonian, $H_{e-i}^{\rm mf}=\mathcal{J}_{zz}V S_z/2$. At voltage $|V|\gg T$ the resonances are well separated since their width is of the order $J^2 V$. The width of the resonance can be estimated more accurately as follows. Let us introduce the eigenbasis of $S_z$, $S_z|m\rangle=m|m\rangle$ with $m=-S,\dots,S$. Neglecting the terms of the second order in $J$ in GKSL equation, we find that the set of eigenvalues $\{l_\alpha\}$ can be approximated as $-i\mathcal{J}_{zz}V(m-m^\prime)/2$, where $m,m^\prime=-S,\dots,S$. The omitted terms can then be taken in account by the first order perturbation theory. Denoting $-\re l_\alpha$ as $1/\tau_{m,m^\prime}$, we obtain
\begin{multline}\label{L19}
\tau^{-1}_{m,m^\prime}=\frac{\pi V g}{4}\biggl\{ q\left[S(S+1)-\frac{m^{2}+m^{\prime 2}}{2}\right]\\+\left(1-q\right)(m-m^\prime)^{2}-\frac{1}{2}q p (m+m^\prime)\biggl\} ,
\end{multline}
where $g=(\mathcal{J}^T\mathcal{J})_{xx}+(\mathcal{J}^T\mathcal{J})_{yy}=\mathcal{J}^2_{xx}+\mathcal{J}^2_{yy}+\mathcal{J}^2_{yx}+\mathcal{J}^2_{zx}+\mathcal{J}^2_{zy}$, and
\begin{align}
    q & = \frac{\mathcal{J}^2_{xx}+\mathcal{J}^2_{yy}+\mathcal{J}^2_{yx}}{\mathcal{J}^2_{xx}+\mathcal{J}^2_{yy}+\mathcal{J}^2_{yx}+\mathcal{J}^2_{zx}+\mathcal{J}^2_{zy}} ,
\notag \\
p & = \frac{2|\mathcal{J}_{xx}\mathcal{J}_{yy}|}{\mathcal{J}^2_{xx}+\mathcal{J}^2_{yy}+\mathcal{J}^2_{yx}} .
\end{align}
The width of the resonance at $\omega=0$ is determined by the set of Korringa rates $1/\tau_{m,m}$. It is worthwhile to mention the relation, $\tau_{m,m}\propto 1/q$, that gives rise to bunching of the backscattering electrons and to unlimited backscattering Fano factor in the limit $q\to 0$ \cite{Kurilovichi2019b}. In this regime the other resonances are much wider. Generically, for not too small $q$, all resonances have widths which are of the same order.

%%%%%%%%%%%%%%%%%%%%

\section{\label{sec:four}
The backscattering current noise 
for $S={1}/{2}$ }

In the case of spin $S=1/2$ it is convenient to perform transformation from the density matrix to the (super)vector with the help of the following parametrization,
\begin{equation}
|\rho^{(\lambda)}_{S}\rangle=\frac{1}{2}\left \{\Tr \rho^{(\lambda)}_{S}, \Tr[\rho^{(\lambda)}_{S} \bm{S}]\right \} .
\end{equation}
In this representation the matrices $\bm{\mathcal{L}}_{0,1,2}$ can be written explicitly. In particular, we find
\begin{equation}
\bm{\mathcal{L}}_{0}=
\frac{\pi T}{2}
\begin{pmatrix}0 & \bm{0}\\
V \mathcal{X}/T & -\Gamma  ,
\end{pmatrix}
\end{equation}
where $3\times 3$ matrix $\Gamma$ is defined as 
\begin{equation}\label{s1f14}
\Gamma_{jk}=\frac{1}{\pi T}\left( \delta_{jk} \eta_{ll} -\frac{\eta_{jk }+\eta_{kj}}{2}+VJ_{i z}\varepsilon_{ijk}\right) .
\end{equation}
We note that at $V=0$ the matrix $\Gamma$ coincides with the matrix $\Gamma_\sigma(\omega=0)$, cf. Eq. \eqref{eq:Gamma:s:def}. The right and left zero eigenvectors of $\bm{\mathcal{L}}_0$ can be written explicitly,
\begin{equation}
|0\rangle=\begin{pmatrix}
1 \\
V \Gamma^{-1}\mathcal{X}/T
\end{pmatrix}, \quad 
\langle \tilde{0}| = \{1,0,0,0\} .
\end{equation}
Next, the matrix $\bm{\mathcal{L}}_{1}$ is given by
\begin{equation}
\bm{\mathcal{L}}_{1}=-\frac{i\pi V}{8}\begin{pmatrix}
g & -\frac{1}{2}\coth\left(\frac{V}{2T}\right)\mathcal{X}^T\\
2\coth\left(\frac{V}{2T}\right)\mathcal{X} & 2 Q 
\end{pmatrix} .
\end{equation}
Here we introduce the symmetric matrix
\begin{equation}\label{s1f15}
Q_{jk}=\mathcal{J}_{jx}\mathcal{J}_{kx}+\mathcal{J}_{jy}\mathcal{J}_{ky}-g\delta_{jk}/2.
\end{equation}
The matrix $\bm{\mathcal{L}}_2$ can be cast in the following form,
\begin{equation}
\bm{\mathcal{L}}_{2}=-\frac{\pi V}{16}\begin{pmatrix}
g\coth\left(\frac{V}{2T}\right) & -\frac{1}{2}\mathcal{X}^T\\
2 \mathcal{X} & 2\coth\left(\frac{V}{2T}\right) Q 
\end{pmatrix} .
\end{equation}
Finally, using Eq. \eqref{gen:noise:f2}, we obtain the expression for the backscattering current noise for spin $S=1/2$,
\begin{align}
%\label{G7}
\mathcal{S}_{\rm bs}(\omega) = & \frac{\pi V^{2}}{16T}\coth\left(\frac{V}{2T}\right)\mathcal{X}^T\left(\mathbf{1}+\left(\frac{2\omega}{\pi T}\right)^{2}{\Gamma}^{-2}\right)^{-1}\notag\\&\times\Biggl [\frac{V}{2T} \left(g-\frac{V}{2T}\coth\left(\frac{V}{2T}\right)\mathcal{X}^T {\Gamma}^{-1}\mathcal{X}\right){\Gamma}^{-2}\notag\\&-\Gamma^{-1}\left (\coth\left(\frac{V}{2T}\right)\mathbf{1}+\frac{V}{T} {Q}{\Gamma}^{-1}\right )\Biggr ] \mathcal{X}\notag\\&+\frac{\pi V}{8}\left(g\coth\left(\frac{V}{2T}\right)-\frac{V}{2T}\mathcal{X}^T {\Gamma}^{-1}\mathcal{X}\right)  .
\label{eq:noise:s12}
\end{align}
We mention that this result for the backscattering current noise generalizes the result found in Ref. \cite{Nagaev2018} to the case of arbitrary ratio between $\mathcal{J}_{zz}$ and the other components of the exchange matrix (see the discussion in Refs.\cite{Kurilovichi2019c,Nagaev2019}).

For a sake of completeness, we present here also the result for the average backscattering current for spin $S=1/2$ \cite{Kurilovichi2019b},
\begin{equation}
I_{\rm bs}=\frac{\pi V}{8}\left(\frac{V}{2T}\coth\left(\frac{V}{2T}\right) \mathcal{X}^T {\Gamma}^{-1}\mathcal{X}-g\right) .
\label{eq:cur:s12}
\end{equation}
As one can easily check from Eqs. \eqref{eq:noise:s12} and \eqref{eq:cur:s12}, the backscattering current noise $\mathcal{S}_{\rm bs}(\omega)=|I_{\rm bs}|$ for $J|V|\ll \omega \ll |V|$  and $\mathcal{S}_{\rm bs}(\omega=0)=2T |I_{\rm bs}/V|$ for $V\ll J T$.

As follows from Eq. \eqref{eq:noise:s12}, the positions and widths of the resonances in $\mathcal{S}_{\rm bs}(\omega)$ are determined by the eigenvalues $ \Omega_{-1,0,1}$ of the matrix $\pi T\Gamma/2$. In the case $|V|\gg T$, they can be found as 
\begin{align}
\Omega_0 &=\frac{i}{\tau_{1/2,1/2}}+\frac{i}{\tau_{-1/2,-1/2}} \equiv \frac{i}{\tau_K} ,\notag \\
\Omega_{\pm 1} &= \pm  \frac{1}{2}\mathcal{J}_{zz}V+\frac{i}{\tau_{1/2,-1/2}} .
\end{align}

To illustrate the origin of the resonances we consider the lower triangular exchange matrix 
with $|\mathcal{J}_{xx}-\mathcal{J}_{yy}|, |\mathcal{J}_{zx}| \ll |\mathcal{J}_{xx}|, |\mathcal{J}_{yy}|,|\mathcal{J}_{zz}|$, and $\mathcal{J}_{yx}=\mathcal{J}_{zy}=0$. Simplifying the general expression \eqref{eq:noise:s12}, we find
\begin{align}
\mathcal{S}_{\rm bs}(\omega) & = 	2\pi V
\Biggl \{\frac{1}{4}(\delta \mathcal{J})^2+
\frac{(\delta \mathcal{J})^2}{1+\left(\omega\tau_{K}\right)^{2}} \notag \\
& +\frac{\mathcal{J}_{zx}^{2}\delta \mathcal{J}}{|\mathcal{J}_{xx}|+|\mathcal{J}_{yy}|}
\sum_{s=\pm}
\frac{1}{1+4 (\omega-s J_{zz}{V}/{2})^{2} \tau_{K}^{2}} 
\Biggr\} ,
\label{G9}
\end{align}
where $\delta\mathcal{J}=|\mathcal{J}_{xx}|-|\mathcal{J}_{yy}|$. As one can see from Eq. \eqref{G9}, the appearance of side resonances at $\omega=\pm \mathcal{J}_{zz}V/2$ requires the presence of two types of asymmetry in the exchange matrix: the diagonal elements $\mathcal{J}_{xx}$ and $\mathcal{J}_{yy}$ should be different and, in addition,  $\mathcal{J}_{zx}$ should be nonzero. Depending on the sign of $\delta\mathcal{J}$ the current noise at  $\omega=\pm \mathcal{J}_{zz}V/2$ can have a maximum or a minimum. 

\begin{figure}[t]
\centerline{\includegraphics[width=0.45\textwidth]{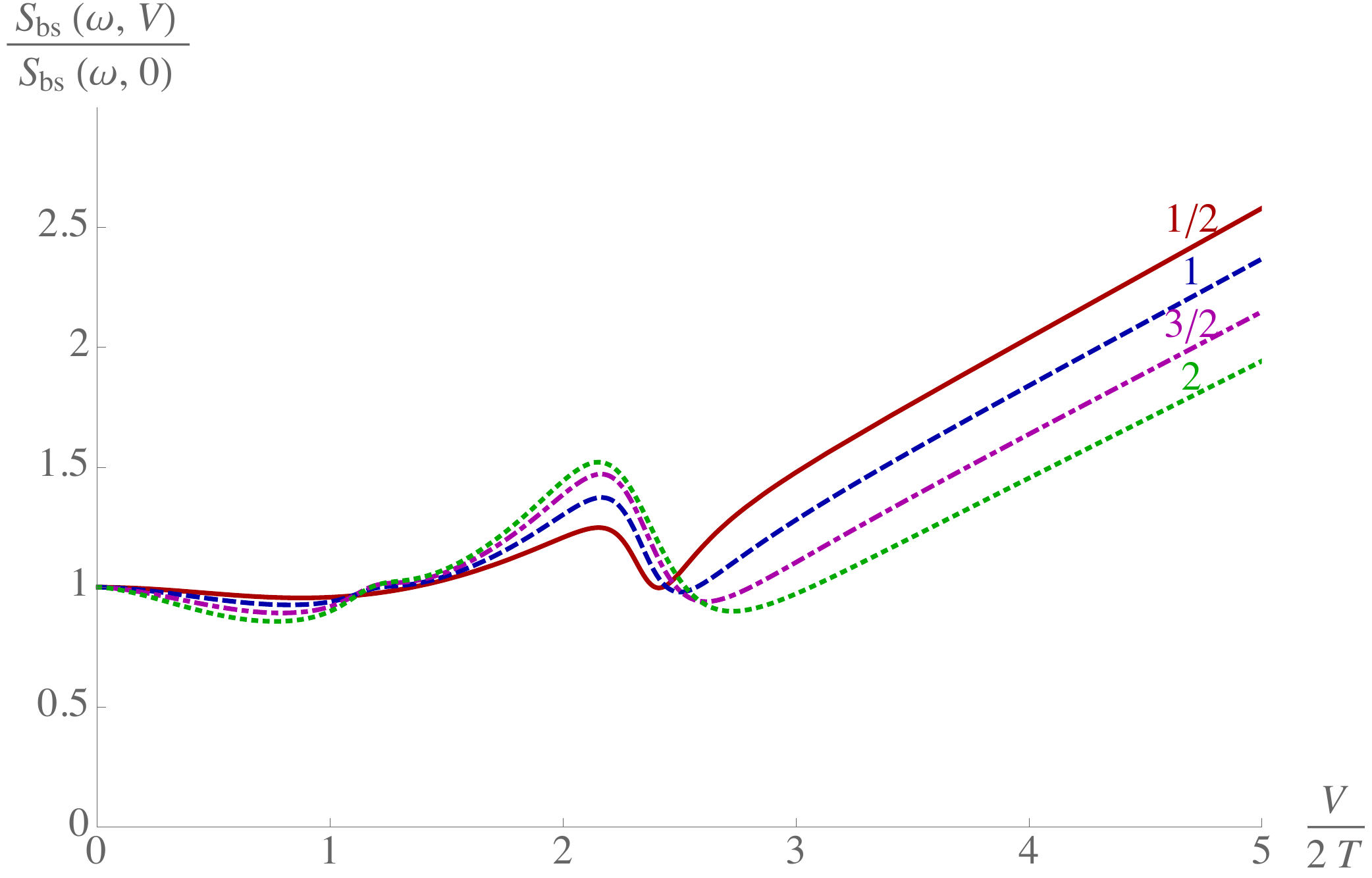}}
\caption{The dependence of the backscattering current noise normalized by its value at zero bias voltage, $\mathcal{S}_{\rm bs}(\omega,V)/\mathcal{S}_{\rm bs}(\omega,0)$, on the dimensionless voltage $V/(2 T)$, for $S=1/2,1,3/2,$ and $2$. The exchange interaction is the same as in Fig. \ref{Figure3}, i.e., $\mathcal{J}_{xx}=\mathcal{J}_{zx}=\mathcal{J}_{zz}=0.01$, $\mathcal{J}_{zy}=\mathcal{J}_{yx}=0.005$, and $\mathcal{J}_{yy}=0.02$. The frequency is equal to $2\omega/(\pi T) = 0.015$.}
\label{Figure4}
\end{figure}

As one can see from Eq. \eqref{G9}, the backscattering current noise has resonances as a function of voltage bias at a fixed frequency. For spin $S>1/2$ the non-monotonic behavior of $\mathcal{S}_{\rm bs}$ with $V$ becomes even more involved as shown in Fig. \ref{Figure4}.  

%%%%%%%%%%%%%%%%%%%%%%%%%%%
%%%%%%%%%%%%%%%%%%%%%%%%%

\section{\label{sec:five}Discussion and conclusion}

\subsection{The effect of the electron-electron repulsion}

In order to take into account the electron-electron repulsion we must use the Luttinger liquid formalism \cite{Giamarchi}. The sole effect of the electron-electron interaction at the helical edge is the modification of the expression for the spin-spin correlation function \cite{PhysRevB.92.045430,PhysRevB.93.241301}. Therefore, the kernel $\Pi_V^{(\lambda)}$, cf. Eq. \eqref{PiV:def}, will be transformed to 
\begin{equation}
\Pi_V^{(\lambda)} =\mathcal{F}_{V,T} \begin{pmatrix}
 f_\lambda^+\left(V/T\right)
& -i f_\lambda^-\left(V/T\right) & 0 \\
i f_\lambda^-\left(V/T\right)& f_\lambda^+\left(V/T\right)& 0 \\
0 & 0& K^{-1} \mathcal{F}_{V,T}^{-1}
\end{pmatrix} .
\end{equation}
Here the Luttinger liquid parameter $K$ is assumed to be within the range $1/2<K \leqslant 1$, corresponding to moderate repulsion ($K=1$ corresponds to the noninteracting case). The function $\mathcal{F}_{V,T}$ is defined as follows
\begin{align}
    \mathcal{F}_{V,T}= & \left(\frac{2\pi Ta}{u}\right)^{2K-2}
   \frac{2T}{V} \sinh\left (\frac{V}{2T}\right )
   \notag\\ 
   & \times 
    B\left(K-i\frac{V}{2\pi T},K+i\frac{V}{2\pi T}\right) .
\end{align}
Here $B(x,y)$ stands for the Euler beta function, $u$ is the renormalized velocity of helical edge states, and $a$ denotes the length scale which corresponds to the ultraviolet cutoff. 

In the case of large bias voltage, $|V|\gg T$, the matrix $\Pi_V^{(\lambda)}$ simplifies to (cf. Eq. \eqref{eq:Pi2}),
\begin{gather}
\Pi_V^{(\lambda)}=  \frac{V}{2\Gamma(2K) T}\left(\frac{aV}{u}\right)^{2K-2}
 e^{- i\lambda \sgn V}
\begin{pmatrix}
1 & -i & 0\\
i & 1 & 0\\
0& 0& 0
\end{pmatrix} .
\end{gather}
In the case of lower triangular exchange matrix $\mathcal{J}_{jk}$, this form of $\Pi_V^{(\lambda)}$ implies the following replacements in the final result \eqref{gen:noise:f2}
for the backscattering current noise at $|V|\gg T$,
\begin{gather}
\mathcal{J}_{jk} 
\rightarrow
\frac{\mathcal{J}_{jk}}{\sqrt{\Gamma(2K)}}\left(\frac{aV}{u}\right)^{K-1} , \quad j=x,y,z, \, k=x,y,
\notag
\\
 \mathcal{J}_{zz} \rightarrow \mathcal{J}_{zz}.
 \label{eq:e-e:rule}
\end{gather}
With this simple prescription for the inclusion of electron-electron repulsion in hands, we are able to make the following predictions. At $|V|\gg T$ the positions of the resonances in $\mathcal{S}_{\rm bs}(\omega)$ are essentially insensitive to the presence of interaction. However, the repulsive electron-electron interaction broadens the resonances. 
In the regime of large frequencies, $J |V| \ll |\omega| \ll |V|$, 
the relation $\mathcal{S}_{\rm bs}(\omega)/ |I_{\rm bs}|\to 1$ survives in the presence of interaction.

At low bias voltage, $|V|\ll J T$,  one can use the results for the backscattering current noise, cf. Eq. \eqref{noise:low}, and for the average backscattering current, cf. Eq. \eqref{eq:Ibs:V0}, provided the following substitutions are performed:
\begin{gather}
\mathcal{J}_{jk} 
\rightarrow
\frac{\mathcal{J}_{jk}\Gamma(K)}{\sqrt{\Gamma(2K)}}\left(\frac{2\pi T a}{u}\right)^{K-1} , \quad j=x,y,z, \, k=x,y,
\notag
\\
 \mathcal{J}_{zz} \rightarrow \mathcal{J}_{zz}/\sqrt{K}.
 \label{eq:e-e:rule0}
\end{gather}
In particular, this implies that the Nyquist-type relation \eqref{eq:Nyquist:0} still holds in the presence of repulsive electron-electron interaction.

\subsection{Dilute magnetic impurities}

Experimentally, the helical edge can be contaminated by many magnetic impurities. Let us assume that they are situated in a $\delta$-layer in a HgTe/CdTe quantum well heterostructure. Then in the limit of  dilute magnetic impurities the average backscattering current and noise are given as sum of independent contributions from each magnetic impurity. Since the exchange interaction $\mathcal{J}_{jk}$ depends exponentially on the distance from the edge \cite{Kurilovichi2017},  significant backscattering is caused by magnetic impurities situated close to the helical edge only. As the consequence, the approximation that all magnetic impurities relevant for backscattering have the same matrix $\mathcal{J}_{jk}$, is well justified. Thus the average backscattering current and noise are proportional to the number of magnetic impurities situated at the helical edge. Dispersion in parameters of the exchange interaction for magnetic impurities at the edge is small and, thus, will only give rise to weak broadening of the resonances.  In the case of magnetic impurities scattered randomly in the direction of quantum well growth, the exchange interaction $\mathcal{J}_{jk}$ for each magnetic impurity situated at the helical edge is also random. Therefore, the average backscattering current and noise for this case can be obtained from averaging of the results for a single impurity over $\mathcal{J}_{jk}$ with a proper distribution. Inevitably, this would lead to a significant broadening of the resonances in the current noise.

Due to the indirect exchange interaction, the spins of magnetic impurities can become correlated, which, in turn, can affect the backscattering events. In the presence of electron-electron interaction the indirect exchange interaction depends on a distance $R$ between impurities as a power law, $(a/R)^{2K-1}$ \cite{LeeLee2015}. Since we assumed  that $K>1/2$, the indirect exchange interaction decays with increase of $R$. On the other hand, for $K<1/2$ the Kondo physics  dominates and magnetic impurities cannot be considered as independent even at distances $R\gg a$ \cite{Yudson2018}. Therefore, in the considered case, $K>1/2$, 
the condition $n_{\rm imp} a\ll 1$, determines the dilute limit for magnetic impurities. 
Here $n_{\rm imp}$ stands for the one-dimensional concentration of  magnetic impurities at the helical edge.
\color{black}

\subsection{Relation to experiments}

As shown in Fig. \ref{Figure4}, we predict a non-monotonic dependence of the backscattering current noise on voltage at fixed frequency. We note that similar non-monotonic dependences of the current noise (but at $\omega=0$) have been observed for a 2D hole system in the regime of hopping conductivity \cite{Kuznetsov2000}, and  for tunneling via interacting pairs of localized states in a 2D electron system \cite{Safonov2003,Savchenko2004}. In the latter case the non-monotonicity in the current noise has been attributed to correlations between tunneling events via different  localized states. We stress that in our case the non-monotonicity of $\mathcal{S}_{\rm bs}$ is due to a mechanism similar to electronic paramagnetic resonance. The bias voltage plays the role of an effective magnetic field that lifts the spin degeneracy for the magnetic impurity. Transport at finite frequency allows to induce transitions between split energy levels of the magnetic impurity.
Then backscattering off the magnetic impurity leads to the Fano-type resonances in the current noise at non-zero frequencies  which are similar to an asymmetric  electronic paramagnetic resonance \cite{Shavit2019a,Shavit2019b}. 
\color{black}
Also $\mathcal{S}_{\rm bs}(\omega)$ can be contrasted with the resonances of the Lorentzian form measured in the frequency dependent shot noise in transport via single electron transition \cite{Delsing}. It is worthwhile to mention that recently the Fano-type resonances in the frequency current noise have been predicted in a double dot Aharonov-Bohm interferometer \cite{PhysRevB.101.235144}. Their origin was an interplay between Coulomb blockade and Rabi interference in the presence of non-zero Aharonov-Bohm flux. 

Recently, the zero-frequency current noise measured by scanning tunneling microscope situated at the top of a magnetic adatom has been demonstrated to be a tool to resolve its energy structure \cite{Burztlaff2015,Pradhan2018,Cocklin2019}. Our results suggest that the frequency resolved backscattering current noise can serve as a sensitive probe to measure various physical characteristics of a dynamical impurity spin. Measurement of the resonant frequencies in the dependence of $\mathcal{S}_{\rm bs}$ on $\omega$ allows one to estimate a value of the dimensional exchange interaction $\mathcal{J}_{zz}$. Measurement of widths of the resonances enables to estimate a magnitude of the other components of the exchange matrix $\mathcal{J}_{jk}$. Observation of more that one non-zero resonant frequency signals that the impurity spin is larger than $1/2$. However, we mention that we are not aware of measurements of the frequency dependence of the current noise at the helical edge in 2D topological insulators to date.

Let us estimate the resonant frequency in the case of 2D topological insulator in a HgTe/CdTe quantum well with Mn$^{2+}$ ions. For a typical bias current, $I \approx 1$ nA we can estimate the bias voltage across the helical edge as $V\sim h I/e^2 \approx 2.6 \cdot 10^{-5}$ V. We note that this voltage corresponds to the temperature of the order of $0.3$ K. Then for  the resonant frequency we find $f \sim e J V/h \approx 6 $ MHz. Here we used that $J \approx 10^{-3}$. The widths of the resonances can be estimated, roughly, as $\Delta f \sim J f \approx 6$ kHz.  

In our theory we neglects the retardation effects. This limits application of our results to not too large frequencies, $\omega\ll v/L$. Here $L$ denotes the length of the helical edge. Estimating velocity of the helical edge as $v\approx 10^5$ m/s \cite{Xiao}, we find $v/L \approx 100$ GHz for $L=1 \ \mu$m.  
\color{black}

In the consideration above we neglected the local anisotropy of the impurity spin Hamiltonian, $\mathcal{D}_{jk} S_j S_k$, that can be present in a real system. We remind that this can be justified for $|\mathcal{D}_{jk}|\ll \max\{J^2 T,|J V|\}$.
The local anisotropy results in splitting of energy levels for the impurity spin. In turn, this affects the average backscattering current \cite{Kurilovichi2019a}. We expect that the modulation of the average backscattering current due to local anisotropy can lead to non-monotonic dependence of the shot noise on a bias voltage already at zero frequency.

\subsection{Conclusion}

To summarize, we have studied  the helical edge coupled to the dilute dynamical magnetic impurities. We considered the case of an arbitrary spin $S$ and a general form of the exchange interaction allowed by the symmetries. Under the assumption of weak exchange interaction, we derived  analytic expressions for the backscattering current noise at finite frequency and studied its dependence on the
temperature and applied bias voltage. Our main finding is that in addition to the Lorentzian resonance at zero frequency the backscattering current noise has additional Fano-type resonances. Such resonant structure of the current noise as a function of frequency transforms into a non-monotonic behaviour of $\mathcal{S}_{\rm bs}$ as a function of the bias voltage. We proposed the backscattering current noise measured at finite frequency as a sensitive tool to access a fine structure of  the impurity spin Hamiltonian.

\begin{acknowledgments}
The authors are grateful to V. Khrapai for very useful comments. The authors are indebted to Y. Gefen, P. Kurilovich and V. Kurilovich for collaboration at the initial stage of this project. The research was partially supported by the Russian Ministry of Science and Higher Education, the Alexander von Humboldt Foundation, the Israel Ministry of Science and Technology (Contract No. 3-12419), the Israel Science Foundation (Grant No. 227-15) and the US-Israel
Binational Science Foundation (Grant No. 2016224). Hospitality by Tel Aviv University, the Weizmann Institute of Science, Landau Institute for Theoretical Physics, and National Research University Higher School of Economics is gratefully acknowledged. M.G. acknowledges a travel grant by the BASIS Foundation (Russia).
\color{black}
\end{acknowledgments}

\bibliography{bib}
\end{document}